\documentclass[conference,letterpaper]{IEEEtran}
\usepackage{fancyhdr}
\usepackage{fancyhdr}

\IEEEoverridecommandlockouts

\usepackage{amssymb}
\usepackage{amsmath}
\usepackage{url}
\usepackage{multirow}
\usepackage{graphicx}

\begin{document}
\title{A Study on Classification in Imbalanced and Partially-Labelled Data Streams}



\author{
    \IEEEauthorblockN{R. J. Lyon, J. M. Brooke, J. D. Knowles}
    \IEEEauthorblockA{
    School of Computer Science\\
	University of Manchester\\
	Manchester, UK\\
    }
    \and
    \IEEEauthorblockN{B. W. Stappers}
    \IEEEauthorblockA{
    Jodrell Bank Centre for Astrophysics\\
	University of Manchester\\
	Manchester, UK\\
    }
    
    \IEEEcompsocitemizethanks{
    \IEEEcompsocthanksitem This work was supported by grant EP/I028099/1 for the University of Manchester Centre for Doctoral Training in Computer Science, from the UK Engineering and Physical Sciences Research Council (EPSRC). \newline 978-1-4799-0652-9/13/\$31.00~\copyright2013 Crown
IEEE}
}

\maketitle
\thispagestyle{plain}

\fancypagestyle{plain}{
\fancyhf{}	
\fancyfoot[L]{}
\fancyfoot[C]{}
\fancyfoot[R]{}
\renewcommand{\headrulewidth}{0pt}
\renewcommand{\footrulewidth}{0pt}
}

\pagestyle{fancy}{
\fancyhf{}
\fancyfoot[R]{}}
\renewcommand{\headrulewidth}{0pt}
\renewcommand{\footrulewidth}{0pt}

\begin{abstract}
The domain of radio astronomy is currently facing significant computational challenges, foremost amongst which are those posed by the development of the world's largest radio telescope, the Square Kilometre Array (SKA). Preliminary specifications for this instrument suggest that the final design will incorporate between 2000 and 3000 individual 15 metre receiving dishes, which together can be expected to produce a data rate of many TB/s. Given such a high data rate, it becomes crucial to consider how this information will be processed and stored to maximise its scientific utility. In this paper, we consider one possible data processing scenario for the SKA, for the purposes of an all-sky pulsar survey. In particular we treat the selection of promising signals from the SKA processing pipeline as a data stream classification problem. We consider the feasibility of classifying signals that arrive via an unlabelled and heavily class imbalanced data stream, using currently available algorithms and frameworks. Our results indicate that existing stream learners exhibit unacceptably low recall on real astronomical data when used in standard configuration; however, good false positive performance and comparable accuracy to static learners, suggests they have definite potential as an on-line solution to this particular big data challenge.
\end{abstract}


%
\IEEEpeerreviewmaketitle

\section{Introduction}
Data streams have arisen naturally from the ever increasing volumes of data being generated by modern computational systems. Typically rapidly generated, temporally ordered, and infeasible to store in their entirety, data streams pose a significant challenge to those seeking to unlock the knowledge they contain. In recent years, considerable research effort has been expended towards identifying and solving these problems, leading to the development of many effective data stream learners. However the main focus of this work has been upon learning from streams that possess a reasonably balanced class distribution, and completely labelled or in some cases partially labelled data. In situations where the class balance is heavily skewed, or where it is unrealistic to expect more than a small fraction of the stream to be labelled, we face a new set of challenges that undermine the effectiveness of existing approaches. In this paper, the results of an empirical investigation are presented which demonstrate how the performance of Hoeffding bound based data stream classifiers, can degrade when faced with an increasingly imbalanced data stream. Motivated by a real world problem faced by the radio astronomy community, this paper seeks to add to data stream research efforts, by drawing increased attention to a learning scenario that is likely to become commonplace as data streams appear in ever more diverse domains.

\section{Motivation}
Radio Pulsars are extremely dense, rapidly rotating stellar remnants formed during the collapse of massive stars. They are a somewhat rare phenomena, which happen to provide a unique environment in which to perform numerous astrophysical experiments \cite{Cordes:2004gi}. These scientific opportunities have motivated the search for pulsars using large radio telescopes. 

In order to find pulsars, the signals arriving at the receiver of a radio telescope must be fed into a computational pipeline, designed to search for periodic broadband radio emission. Those periodic signals possessing a signal-to-noise ratio above a predefined threshold value (based on domain knowledge), are considered to be pulsar `candidates' worthy of further investigation. The search pipeline itself is made up of many processing components. Broadly speaking these can be separated into those that act as data processors which clean and correct the data (e.g. excision of radio frequency interference) and search routines tasked with isolating signals of interest. Crucially the volume of data moving through such search pipelines has been increasing steadily for some time. Indeed as observed in \cite{BatesPhD:1}, pulsar survey data capture rates have exhibited a level of growth closely describable by Moore's Law. This growth reflects advances in technology that have enabled various improvements in survey specifications, from increased survey sampling rates to finer frequency resolution. However there is an inverse relationship between the sampling rate of a pulsar survey and the data capture rate. As the sampling period decreases (thereby increasing sensitivity to shorter period emissions) the data capture rate increases. When coupled with finer frequency resolution and wider observational bandwidths, this relationship creates successive data storage and processing challenges. Similarly an inverse relation exists between the sampling rate of a survey, and the number of signals meeting the candidate selection criteria. As the majority of signals captured at the receiver of a telescope are attributable to noise or interference; nearly the entire set of selected candidates are of no scientific value. Separating those candidates which are likely to lead to discovery from those which are spurious is therefore becoming increasingly difficult to do.

\subsection{The Square Kilometre Array}
The next generation of radio telescope, the Square Kilometre Array (SKA) \cite{SKA1} is currently under development by an international consortium. Due to begin science operations in the next decade, the SKA will produce data at an unprecedented rate, firmly ushering in the era of exascale computing. For pulsar survey operations alone the data rate is predicted to be between $0.43-1.45$ TB/s \cite{Smits:2009dc}. This is many orders of magnitude greater than previous surveys. Using existing technology, it is practically impossible to acquire the hardware and supporting infrastructure required for storing all this data permanently due to financial restrictions. Hence future SKA surveys must either a) utilise technological advances that make it feasible to store such large quantities of data at low cost, or b) process the data in real-time, storing only a fraction of the data for off-line analysis. Although one cannot say with certainty, it appears that real-time analysis is the only tractable and cost effective long term solution to this problem \cite{Smits:2009dc}. If this is indeed the case, then in the future candidate selection will have to be performed on-line, and in real-time. In the absence of an existing real-time selection system or the means to deal with the impending increase in candidates; the possibility exists that important discoveries could be delayed or possibly missed due to inadequate supporting computational infrastructure. Given that present day searches for radio transients\footnote{Typically transients are short non-repeating bursts of high energy radio emission.} are already being conducted in real time \cite{Macquart:2010ko}, it seems likely that the real-time scenario will win out. 

\section{Problem Definition}
\label{sec:one}
We consider the selection of candidate pulsars as a two-class data stream classification problem. We define the completely unlabelled input stream $S={ \lbrace (x_{i}),...,(x_{n}),... \rbrace}$, $i=1,...,n$ as the stream of candidates emerging from a pulsar search pipeline, under a discrete time model. Each candidate in the stream $x_{i} \in X$ is defined as $x_i = \lbrace (y_{i}^{1}),...,(y_{i}^{m}) \rbrace$, where each $y_{i}^{j} \in \mathbb{R} $ for $j=1,...,m$ is a summary statistic or measure that describes the candidate $x_{i}$. We specify binary candidate labels $L = \lbrace 0,1 \rbrace $, where $l$ is an individual label such that $l \in L$ (i.e. uninteresting = 0, interesting = 1). Our goal is to learn a function $f: X \mapsto L $ which maps each candidate to its correct label producing the set of labelled candidates $C = \lbrace (x_{i},f(x_{i})),...,(x_{n},f(x_{n})) \rbrace $. From $C$ the set of positively labelled candidates (those instances with $l=1$) can be obtained, which should be recommended for inspection. As the stream is completely unlabelled, only those candidates selected are ever likely to receive their correct labels (with the remainder necessarily discarded). Even then only a fraction of these may receive labels given the candidate volume, and certainly only after a substantial delay. Thus this scenario can be thought of in terms of two distinct streams. The first from the telescope, the second a biased feedback stream containing data labelled by experts. Crucially the primary stream emerging from the telescope will be heavily imbalanced in favour of the negative class. The Parkes Multibeam Pulsar Survey (PMPS) for example has so far yielded 742 pulsar discoveries \cite{Lorimer:2006ha}, from $\sim 8$ million candidates \cite{Eatough:2010uz}; giving rise to a class balance of roughly +1 : -11,000. If we make the conservative assumption that the class balance of the PMPS is maintained by the SKA, then we can expect to collect some $\sim$200,000,000 negative candidates (given that current estimates suggest the SKA will detect approximately 20,000 new pulsars in total \cite{Smits:2009dc} ). A hypothetical classifier with perfect recall (\ref{eq:recall}) and an accuracy (\ref{eq:accuracy}) of .99 applied to these SKA candidates, would still select 2 million for further analysis. 


\small
\begin{equation} \label{eq:recall}
\textrm{ Recall } = \frac{\textrm{True Positives}}{\textrm{True Positives + False Negatives}}
\end{equation}
\begin{equation} \label{eq:accuracy}
\textrm{Accuracy} = \frac{\textrm{True Positives + True Negatives}}{\textrm{Positives + Negatives}}
\end{equation}
\begin{equation} \label{eq:precision}
\textrm{Precision} = \frac{\textrm{True Positives}}{\textrm{True Positives + False Positives}}
\end{equation}
\begin{equation} \label{eq:fprate}
\textrm{False Positive Rate} = \frac{\textrm{False Positives}}{\textrm{False Positives + True Negatives}}
\end{equation}
\normalsize

This is far too  many for an expert to analyse, and infeasible to follow up on given the costs associated with telescope time. An ideal classifier must therefore maintain a high rate of recall (as missing positive instances is acutely costly given their rarity) and significantly reduce the false positive rate (\ref{eq:fprate}). We summarise the five key data mining challenges for our candidate classifier as follows:
\begin{enumerate}
\item \textbf{Imbalanced class distribution}: the ratio of positive to negative instances will be greatly imbalanced.
\item \textbf{Nonstationarity} : Although the negative class is comprised primarily of noise, various types of interference will be subject to gradual changes over long time scales, and abrupt changes over short ones.
\item \textbf{Unlabelled data}: the data in the stream is completely unlabelled. Drift detection must be performed on unlabelled data, without the prospect of obtaining correct labels within a reasonable time. 
\item \textbf{Real-time processing}: The SKA will produce a high throughput data stream. It is unrealistic for much more than a small sample of this data to be stored, hence processing must likely be done in real-time. 
\end{enumerate}

\section{Related Work}
\subsection{Candidate Selection}
Candidate selection has typically involved the expert selection of candidate signals via some form of summary user interface. During the Swinburne Intermediate Latitude Pulsar Survey \cite{Edwards:2001tm} and PMPS \cite{Manchester:2001fo}, graphical tools were developed which allowed large numbers of candidates to be filtered. A later reprocessing of the PMPS spawned the more sophisticated \textsc{reaper} tool, which enabled candidates to be viewed via a customizable graphical plot \cite{Faulkner:2004cl} ultimately leading to the discovery of 128 pulsars. A new version of this tool called \textsc{jreaper} was then developed and led to the discovery of 28 pulsars \cite{Keith:2009jo}. More recently machine learning techniques have been used to filter candidates. In \cite{Eatough:2010uz} a multi-layered perceptron (MLP) was used to perform an automated re-analysis of a data sample taken from the PMPS. The implementation was capable of recalling up to 93\% of the pulsars present in PMPS test samples. Other methods utilised include Gaussian Mixture models as applied in \cite{lee:2012:kramer}, to rank pulsar candidates from the Fermi 2FGL catalogue.
\subsection{Imbalanced \& Unlabelled Streams}
The problem of class imbalance is usually tackled by either assigning different costs to training examples thereby reweighting the balance, or by resampling the original dataset to achieve a desired balance. Resampling can be done in many ways, either by over-sampling the minority class or under-sampling the majority class \cite{Chawla:2002:bk}. The SMOTE algorithm developed by Chawla et al. \cite{Chawla:2002:bk} mixes both of these approaches. Empirical results suggest that SMOTE can successfully improve the accuracy of classifiers such as C4.5 and Na\"{\i}ve Bayes on the minority class. In \cite{Haibo:2011:cs} Chen et al. develop their recursive weighted ensemble approach (REA) for classifying nonstationary imbalanced data streams. REA adaptively pushes minority class examples into the current data chunk to balance the class distribution. An in-depth discussion of the problems associated with learning from imbalanced data can be found in \cite{Haibo:2009:fb}. In terms of unlabelled streams, the Semi-supervised learning paradigm \cite{Mitchell:1997ua} has typically been applied when large data sets are continually produced or when the labelling of the data is prohibitively costly \cite{Vittaut:2002kd}. Some representative works in this area include: the decision tree based algorithm for classifying concept drifting and unlabelled data streams called SUN developed in \cite{Peipei:2010:xh,Wu:2012ti}; the relational K-means based transfer semi-supervised SVM (RK-T$S^{3}$VM) developed for classifying unlabelled drifting data streams \cite{Zhang:2009:ji}, and the CSL-Stream algorithm (Concurrent Learning of Data Streams) that clusters and classifies data at the same time \cite{Nguyen:2011:ng}.
\section{Experiments}
 Amongst the most successful of the single model data stream classifiers has been the very fast decision tree (VFDT) \cite{Hulten:2001:MTD}. VFDT is an incremental any time decision tree learner that uses constant memory, and permits model updates in time proportional to the tree depth and data dimensionality. VFDTs low runtime and space complexity therefore make it ideal for use on data streams, particularly as VFDT's can be dynamically pruned if required to operate under strict memory constraints. The output of a VFDT is also guaranteed to be asymptotically similar to a conventional learning algorithm.  VFDT achieves this by using the Hoeffding bound (\ref{eq:hoeffding}) to choose with a high probability, those split attributes that would have been selected given access to all the data as in the non-streamed scenario. By calculating the observed mean $\bar{r}$ of an attribute, the bound is able to determine with confidence $1-\delta$ (where $\delta$ is user supplied), that the true mean of an attribute $r$ is at least $\bar{r}-\epsilon$ where,
 
\small
\begin{equation}\label{eq:hoeffding}
\epsilon = \sqrt{\frac{R^{2} \textrm{ }ln(1 / \delta)}{2n}}\textrm{.}
\end{equation}
\normalsize

Although not intended to operate on unlabelled data, we begin looking at ways to solve the candidate selection problem by testing the effectiveness of existing algorithms such as the VFDT. Using the MOA stream mining framework \cite{moa}, we test an implementation of the VFDT called the \textit{HoeffdingTree} \cite{moa} on imbalanced and unlabelled streams. We compare the performance of the \textit{HoeffdingTree} to three static classifiers including a standard decision tree, a SVM and Na\"{\i}ve Bayes. Our tests were designed to reveal how well the \textit{HoeffdingTree} performs if i) made to learn from only the examples in the stream and ii) pre-trained before being taken on-line. Tests of other VFDT tree based algorithms including the \textit{HoeffdingOptionTree \cite{Pfahringer:2007:gh}}, \textit{AdaHoeffdingOptionTree}\footnote{This is a \textit{HoeffdingOptionTree} with the leaves modified so they store an estimation of the current error.} , and finally \textit{OzaBag}\cite{Oza:2005:nc} and \textit{OzaBoost}\cite{Oza:2005:nc} both using the \textit{HoeffdingTree} have also been completed (\textit{in preparation}).  In each case the parameters used for these algorithms were set to the defaults provided by MOA.
\subsection{Data}
\begin{table}
\center
\tabcolsep=0.11cm
    \begin{tabular}{|l|l|l|l|}
    \hline
    Dataset & Instances & Attributes / Type  & Balance \\ \hline
    Pulsar & 11,219,848 & 22 / Continuous, numerical & +1 : -7500 \\
    Skin & 245,057 & 3 ~/ Discrete, numerical& +1 : -5 \\
    MiniBoone & 130,065 & 50 / Continuous, numerical& +2 : -5 \\
    Magic & 19,020 & 10 / Continuous, numerical& +1 : -2\\ \hline
    \end{tabular}
\caption[Data]{Characteristics of the data sets used.}
\vspace{-2.5em}
\label{tab:data}
\end{table}
In order to test the various data stream classifiers, four datasets were utilised (see Table \ref{tab:data}). The first and largest dataset consisted of pulsar candidates obtained during the High Time Resolution Universe Survey (HTRU) \cite{Keith:2010bl}\footnote{This data is currently not publicly accessible.}. Candidates are represented by 22 continuous numerical attributes, each of which is a summary statistic that describes it in some way. The data set contains only 1,611 positive and 2,593 negative examples correctly labelled by human annotators. The remainder of the dataset is na\"{\i}vely assumed to be negative. However we can constrain the number of actual positives incorrectly labelled in this dataset based on Monte Carlo simulations undertaken by Keith \cite{Keith:2010bl} and later by Levin \cite{LevinPhD:1}, to some small fraction of approximately 879-916 instances. Also of the known 1,108 pulsars in the survey region, 725 have been re-detected in this data. Thus together the number of re-detections and expected discoveries falls within the range of the 1,611 certain positive examples already labelled. The remaining datasets used included the Skin Segmentation, MiniBoone and the Magic Gamma Telescope datasets. These are obtainable from the UCI machine learning repository \cite{UCI}.
\subsection{Methodology}
\begin{figure}[htb]
	\centering
		\includegraphics[keepaspectratio,scale=0.45]{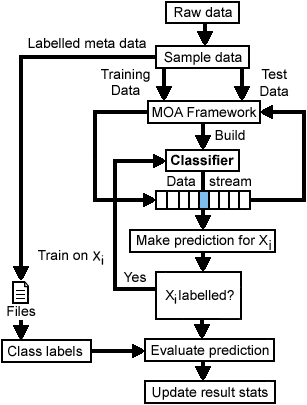}
		\caption[]{An overview of the test framework.}
		\label{fig:1}
		\vspace{-1em}
\end{figure}

Three distinct types of test were undertaken for this work. The first established a baseline level of performance on our data, using static classifiers under a traditional static learning scenario. Here performance on our data sets was assessed using all of the data and stratified 5-fold cross validation (train on 4 folds, test on 1). To test the algorithms under the two streaming scenarios on the other hand (no pre-training vs. pre-trained), the data sets were firstly shuffled (except for the temporally ordered pulsar data) and then randomly sampled in order to generate different levels of class imbalance \footnote{The levels used were: $1.0,0.5,0.3,0.25,0.2,0.1,0.04,0.02,0.01,0.008,$ $0.004,0.002,0.001,0.0004,0.0002,0.00013,0.0001$ (e.g. 0.1 = +1 : -10).}, whilst also varying the proportion of instances labelled \footnote{The labelling proportions used: 0\%,1\%,5\%,10\%,25\%,50\%,75\%,99\% and 100\%.}. The no pre-training scenario tested only the stream classifiers under a test then train model (test on an instance, then train on an instance), such that each of the sampling permutations was treated as a single stream of data. Each test was repeated ten times for a given balance and labelling, allowing results to be averaged over multiple runs. For the pre-trained scenario, the same sampling and test procedure was used except that training and test sets were generated. The learners were then trained prior to being taken on-line for testing following the same test then train approach. All training sets were uniform in size and configuration, containing 200 positive and 1000 negative instances (+1 : -5 balance). The class distribution of these training sets is similar to the test distribution of three out of our four data sets (see Table \ref{tab:data}), the exception being the pulsar data which has a very different test distribution. For the pre-trained tests both static and stream classifiers were used, allowing for a comparison between the performance of static and streamed learners trained on the same small sample of data. The test framework which executed these tests is summarised in Figure \ref{fig:1}. As we assume a discrete time model, for each instance $x_{i}$ arriving at time step $t$, a class prediction is made at time step $t+1$. The learner is trained at time step $t+2$ only if the instance was labelled. Following each prediction a count of true positives, false positives, false negatives and true negatives is updated, by checking the prediction against the correct label of $x_i$. If $x_i$ was unlabelled in the stream, then the correct label is obtained from a meta data file, and then compared to the prediction as before. Thus each prediction is evaluated during testing. However overall statistics are not computed until the end of an individual test run.

Evaluating classifier performance on imbalanced data is known to be difficult, particularly as many metrics are sensitive to the underlying class distribution \cite{Haibo:2009:fb}. We therefore keep track of multiple assessment metrics which include amongst others the F-Score (\ref{eq:fscore}) and the G-Mean \cite{Haibo:2009:fb} (\ref{eq:gmean}). The F-Score is sensitive to the class distribution given its dependence on precision (\ref{eq:precision}). Thus when the data set is large and imbalanced, the numerator of the F-Score equation remains small while the denominator becomes increasingly large. Thus low precision can cause the F-Score to obtain a small value even given a high level of recall (e.g. with recall = .99 and precision = .01, the F1 = .0099) . As the G-Mean is not sensitive to the class distribution in the same way, we present the two together to\newpage provide a representative impression of performance.
\small
\begin{equation} \label{eq:fscore}
\textrm{ F-Score} = 2 \times \frac{\textrm{ Precision } \times \textrm{ Recall }}{\textrm{ Precision } + \textrm{ Recall }}
\end{equation}
\begin{equation} \label{eq:gmean}
\textrm{G-Mean} = \sqrt{\frac{TP}{TP+FN}\times \frac{TN}{TN+FP}}
\end{equation}
\normalsize
\subsection{Results}
In Table \ref{tab:accuracy} baseline accuracy results for static classifiers tested in a traditional learning scenario, are compared to those achieved by the stream classifiers under some representative imbalanced class scenarios. Static learners perform well on our data sets, achieving higher levels of accuracy than the stream learners when the class distribution is reasonably balanced. As the class imbalance increases however, the accuracy of the stream learners tends towards 1, surpassing the accuracy levels achieved by the static classifiers. This increase in accuracy becomes statistically significant for all data sets (using a significance level of 0.05) when the imbalance reaches $\sim 1+ : - 125$. Accuracy tending towards 1 in this manner, appears to happen due to the presence of Na\"{\i}ve Bayes classifiers at each leaf of the \textit{Hoeffding tree}. As each leaf contains a count of those positive and negative examples arriving there, a fully labelled data stream which is heavily imbalanced, will cause the count of negatives at these leaves to increase disproportionately to the positives. This skews the Na\"{\i}ve Bayes predictions towards the negative class. As most examples in the stream are negative, by consistently predicting negative in this way the \textit{Hoeffding tree} is effectively optimising classifier accuracy.
\begin{table}[htdp]
\center
\tabcolsep=0.1cm
    \begin{tabular}{|l|l|l|l|l|l|}
    \hline
    & \multicolumn{5}{c|}{\textbf{Stream Accuracy}}\\ \hline
    \textbf{Balance} & 1.0 & 0.1 & 0.01 & 0.001 & 0.0001 \\ \hline\hline
    Pulsar        & .9174 & .9759 & .9949 & .9992 & .9999 \\ \hline
    Skin           & .9892 & .9860 & .9940 & .9987 & .9996 \\ \hline
    MiniBoone & .9949 & .9995 & 0.9996 & .9999 & .9998\\ \hline
    Magic        & .7154 & .9083 & .9897 & .9990 & - \\ \hline
    \end{tabular}
     \begin{tabular}{|l|l|l|}
    \hline
    \multicolumn{3}{|c|}{\textbf{Static Accuracy}}\\ \hline
    DT & SVM & NB\\ \hline\hline
    .9999 & .9999* & .9676\\ \hline
    .9992 & .9291 & .9239\\ \hline
    .9999 & .9279 & .2827\\ \hline
    .8503 & .7916 & .7269\\ \hline
    \end{tabular}
\caption[Results]{Accuracy of the \textit{HoeffdingTree} without pre-training on a 100\% labelled data stream, versus accuracy of static classifiers trained and tested using stratified 5-fold cross validation (all data). * Here a 90\% train, 10\% test strategy was used.}
\vspace{-2em}
\label{tab:accuracy}
\end{table}
 
\begin{table}[htdp]
\center
\tabcolsep=0.1cm
    \begin{tabular}{|l|l|l|l|l|l|l|}
    \hline
    & \multicolumn{6}{c|}{\textbf{Recall}} \\ \hline
    \textbf{Balance} & 1.0 & 0.5 & 0.1 & 0.01 & 0.001 & 0.0001 \\ \hline\hline
    Pulsar    & .86/.87 & .83/.86 & .81/.82 & .74/.74 & .64/.61 & .06/.04 \\ \hline
    Skin      & .99/ .99 & .98/.98 & .93/.93 & .45/.32 & .08/.07 & .03/0 \\ \hline
    MiniBoone & .98/.99 & .99/.99 & .98/.96 & .96/.84 & .74/.46 & .73/.02 \\ \hline
    Magic & .84/.53 & .13/.31 & .008/.008 & .007/0 & 0/0 & - \\ \hline
    \end{tabular}
\caption[Results]{Recall results for the \textit{HoeffdingTree} with and without pre-training (with/without), on a 50\% labelled data stream.}
\vspace{-2em}
\label{tab:recall}
\end{table}

\begin{figure}[htb]
	\centering
		\includegraphics[keepaspectratio,scale=0.5]{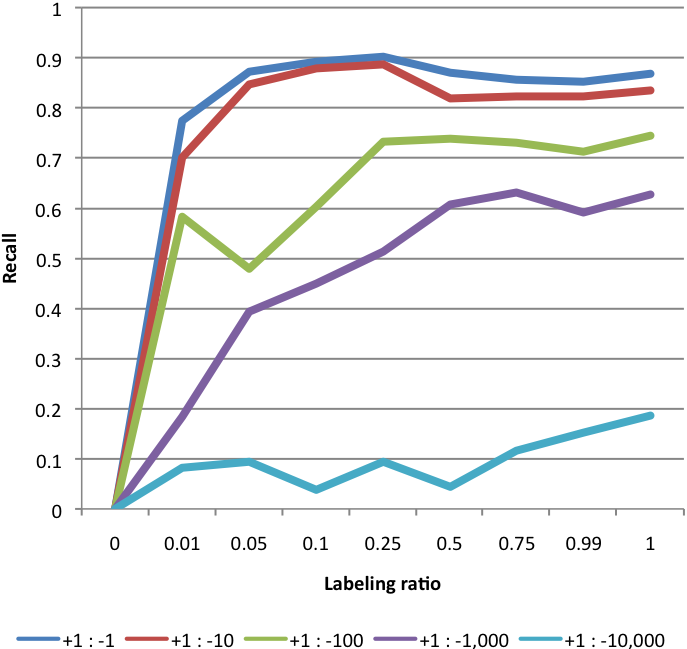}
		\caption[]{Recall rates of the non pre-trained \textit{Hoeffding tree} on pulsar data as the labelling and class balance is altered.}
		\label{fig:2}
		\vspace{-1.0em}
\end{figure}

However the improvement in accuracy comes at the expense of reduced recall. Across all of the datasets tested, recall rates consistently dropped as the class imbalance increased. This is a side effect of almost always predicating the negative class. This effect can be seen clearly in Table \ref{tab:recall}. Even when 50\% of the stream is labelled (which is unrealistic for a SKA scenario) recall rates tend toward zero which is unacceptable for our problem domain. Figure \ref{fig:2} also shows the same drop in recall rates observed when testing a non pre-trained \textit{Hoeffding tree} on pulsar data. In this case recall rates are zero when no labels are available, and slowly increase as more labelled data appears in the stream. Static classifiers trained on only a small sample of data (200 positive and 1000 negative instances) achieved higher recall rates when treating the complete pulsar data stream as a static data set. Recall rates where $\sim .91$ for C4.5 and $\sim .88$ for the SVM respectively. Though both of these static classifiers return far too many false positives. C4.5 returned on average $\sim 138,000$ and the SVM $\sim87,000$.

Compared to a static classifier, the \textit{Hoeffding tree} has a very low false positive return rate when the stream is completely labelled. The previously described preference for applying negative classifications when the stream is imbalanced, means that few positive predictions are ever made. Thus false positives become extremely rare. For the most imbalanced streams only $\frac{1}{37000}$ instances became a false positive. As the proportion of labelled items in the stream is reduced, the false positive rate increases - slightly for the non pre-trained classifiers, but greatly for those pre-trained. This can be seen in Table \ref{tab:recall2} which shows the false positive rates obtained on unlabelled streams. In row one a false positive return rate of .02 equates to $\sim 225,000$ false positives, much worse than the static classifiers. It appears that without labels to learn from, the \textit{Hoeffding tree} maintains an imprecise model similar to the static classifiers, as both are trained on the same sample of data.

\begin{table}[htdp]
 \tabcolsep=0.1cm
 \center
    \begin{tabular}{|l|l|l|l|l|l|l|}
    \hline
    & \multicolumn{6}{c|}{\textbf{False positive rate}} \\ \hline
    \textbf{Balance} & 1.0 & 0.5 & 0.1 & 0.01 & 0.001 & 0.0001 \\ \hline\hline
    Pulsar    & .03 & .02 & .01 & .01 & .03 & .02 \\ \hline
    Skin      & .02 & .02 & .02 & .03 & .02 & .02 \\ \hline
    MiniBoone & .01 & .005 & .02 & .004 & .004 & .1 \\ \hline
    Magic & 0 & 0 & 0 & 0 & 0 & - \\ \hline
    \end{tabular}
\caption[Results]{False positive rates for the \textit{HoeffdingTree} on unlabelled streams, trained on 200 positive \& 1000 negative instances.}
\vspace{-2em}
\label{tab:recall2} 
\end{table}
In comparing the pre-trained and non pre-trained stream classifiers, a statistically significant difference in accuracy was observed. The results are summarised in Tables \ref{tab:resultsnopretrain} and \ref{tab:results}. Non-pre-trained classifiers cannot build a model from an unlabelled stream, thus their accuracy for these streams is zero. On the remaining streams, the pre-trained classifiers initially outperform the non pre-trained, particularly in terms of recall. However as the the size of the data sets and the proportion of labelling increases, the non-pre-trained classifiers achieve higher levels of accuracy, once again by favouring negative classifications. Despite this difference in accuracy, the pre-trained classifiers actually maintain higher recall rates throughout. Though asymptotically their performance does appear to approach that of the non pre-trained classifiers as the number of instances classified increases. Thus it appears that pre-training a classifier only delays the inevitable favouring of the negative class if a large class imbalance exists. In summary these results show that those stream learners we tested, optimise accuracy on imbalanced data sets by almost always predicting the majority class. The net effect of this preference for the majority negative class in our case is to reduce recall and false positive rates, given that the minority class is very rarely predicted. Pre-training a classifier off-line on a sample of training data does in the short term improve recall rates, however this effect is only short lived. 

\begin{table*}[htdp]
\center
\tabcolsep=0.1cm
    \begin{tabular}{|l|l|l|l|l|l|l|l|l|l|l|l|l|l|l|l|l|}
    \hline
    Balance   & \multicolumn{4}{c|}{+1 : -10} & \multicolumn{4}{c|}{+1 : -100}  & \multicolumn{4}{c|}{+1 : -1,000}  & \multicolumn{4}{c|}{ +1 : -10,000}  \\ \cline{1-17}
    Labelling (\%) & 0 & 50 & 75 & 100 & 0 & 50 & 75 & 100 & 0 & 50 & 75 & 100 & 0 & 50 & 75 & 100 \\ \hline\hline
    
    Pulsar    & 0/0 & .9/.87 & .9/.86 & .91/.97 & 0/0 & .86/.76 & .85/.75 & .86/.76 & 0/0 & .78/.59 & .79/.6 & .79/.63 & 0/0 & .29/.14 & .37/.22 & .42/.26 \\
    
    Skin      & 0/0 & .96/.90 & .96/.92 & .97/.93 & 0/0 & .56/.45 & .62/.52 & .68/.60 & 0/0 & .22/.13 & .21/ .12 & .23/.10 & 0/0 & 0/0 & .02/0.1 & .07/0.1 \\
    
    MiniBoone & 0/0 & .98/.98 & .99/.99 & .99/.99 & 0/0 & .92/.9 & .93/.92 & .95/.95 & 0/0 & .67/.56 & .77/.69 & .75/.69 & 0/0 & .05/.04 & .05/0.4 & 0/0 \\
    
    Magic     & 0/0 & .07/.02 & .07/.02 & .07/.02 & 0/0 & .05/.03 & .04/.03 & .07/.03 & 0/0 & 0/0 & 0/0 & 0/0 & - & - & - & - \\
    
    \hline
    \end{tabular}
    \caption[Results]{G-Mean/F1 Score results for the \textit{HoeffdingTree} classifier on the test datasets without pre-training.}
    \vspace{-2.3em}
\label{tab:resultsnopretrain}
\end{table*}

\begin{table*}[htdp]
\center
\tabcolsep=0.1cm
    \begin{tabular}{|l|l|l|l|l|l|l|l|l|l|l|l|l|l|l|l|l|}
    \hline
    Balance   & \multicolumn{4}{c|}{+1 : -10} & \multicolumn{4}{c|}{+1 : -100}  & \multicolumn{4}{c|}{+1 : -1,000}  & \multicolumn{4}{c|}{ +1 : -10,000}  \\ \cline{1-17}
    Labelling (\%) & 0 & 50 & 75 & 100 & 0 & 50 & 75 & 100 & 0 & 50 & 75 & 100 & 0 & 50 & 75 & 100 \\ \hline\hline
    
    Pulsar    & .92/.87 & .9/.86 & .89/.85 & .92/.87 & .92/.57 & .86/.77 & .86/.75 & .88/.75 & .92/.1 & .8/.6 & .77/.6 & .8/.61 & .92/.02 & .23/.08 & .29/.12 & .32/.15 \\
    
    Skin      & .9/.83 & .96/.89 & .97/.89 & .97/.93 & .9/.42 & .67/.55 & .7/.58 & .73/.64 & .91/.09 & .28/.01 & .24/.08 & .2/.07 & .91/.01 & .11/.01 & .13/.03 & .12/.04  \\
    
    MiniBoone & .79/.94 & .99/.99 & .99/.99 & .99/.99 & .76/.83 & .98/.97 & .98/.97 & .98/.98 & .59/.54 & .83/.76 & .95/.91 & .9/.85 & .84/.56 & .83/.74 & .96/.92 & .71/.72 \\
    
    Magic     & 0/0 & .06/.02 & .05/.01 & .1/.04 & 0/0 & 0/0 & .01/.03 & .01/.03 & 0/0 & 0/0 & 0/0 & 0/0 & - & - & - & - \\
    \hline
    \end{tabular}
    \caption[Results]{G-Mean/F1 Score results for the \textit{HoeffdingTree} classifier when trained before classifying the stream.}
    \vspace{-3em}
\label{tab:results}
\end{table*}

\section{Conclusion}
In this paper we have found that the recall capabilities of the \textit{Hoeffding tree} classifier, deteriorate when faced heavily class imbalanced data streams. The imbalance appears to skew the class predictions of the Na\"{\i}ve Bayes classifiers at the leaves of the tree, toward the majority class. This has the effect of improving classifier accuracy, whilst also degrading recall. We have found that pre-training a classifier before taking it on-line does initially improve the recall rate, however the class imbalance makes this effect inherently short term. Although these result suggests that VFDT based classifiers such as the \textit{Hoeffding tree} are unsuitable for solving the candidate selection problem, their capacity to maintain low false positive return rates makes them appealing, particularly as we explore ways in which the VFDT approach could be modified to accommodate imbalanced streams and thereby increase the recall rate. Future work will expand on the results of this investigation, and test other data stream classifiers not exclusively based on the Hoeffding bound. We also intend to analyse the attributes that describe a candidate pulsar, with the aim of removing redundant features which may improve classifier accuracy, and reduce the dimensionality of the data.


\section*{Acknowledgment}
Computational resources were provided by the Jodrell Bank Centre for Astrophysics (JBCA), which has support from the UK Science and Technology Facilities Council (STFC). Experiments were carried on upon data obtained by the Parkes Observatory, which is funded by the Commonwealth of Australia and managed by the Commonwealth Scientific and Industrial Research Organisation (CSIRO). We would also like to thank Dan Thornton (JBCA) for his help in obtaining the HTRU survey data.



%

\end{document}